# The "boson polynomials" of Gel'fand basis and the analytic expression of Wigner's coefficients with multiplicity for the canonical basis of unitary groups

M. Hage-Hassan
Université Libanaise, Faculté des Sciences Section (1)
Hadath-Beyrouth

## Abstract

In this paper we present the generating function method for the derivation of bosons polynomials of Gel'fand basis and Wigner coefficients for the canonical basis of SU(n). We find a new analytic polynomial basis of SU(4) with the exact number of summations, five only. We find also a new algebraic expression of Wigner coefficient with multiplicity for the canonical basis and the isoscalors factors of SU (3) with only three summations.

## 1. Introduction

The theory of unitary groups is of great interest in quantum physics, nuclear and elementary particle. The study of these groups was started in mathematics and several methods have been proposed: the infinitesimal method developed by Shur, Cartan, Killing, Weyl, etc. .., and the Weyl global method [1-10] whose starting point the matrix elements of SU(n). Weyl find the connection between the representation of the symmetric group and the unitary group. Weyl also find the basis vectors of the irreducible representation labeled by the highest weights $[h]_n = [h_{1n}, h_{n-1},..., h_{nn}]$ and the dimension formula. The reduction of the representation with highest weight $[h]_n$ of U(n) to U(n-1) with highest weight $[h]_{n-1}$ is given in terms of Weyl branching law.

$$h_{1,n} \geq h_{1,n-1} \geq h_{2,n} \geq h_{2,n-1} ... \geq h_{n-1,n-1} \geq h_{nn}]$$

Using the "Weyl's branching law" Gelfand-Zetlin introduce the basis of representation of U (n), function of $n(n+1)/2$ indices, and later proved the orthogonality of this basis. Moreover, Cartan find that these irreducible representations are polynomials of the fundamental representations $[1,..,0],...[1,..,1]$, whose number is $2^n - 1$.

In physics the Schwinger's method [11] of bosons calculus, has been extended to study the homogenous polynomials basis for the irreducible representation of U(n) by Bargmann and Moshinsky and other [12-18]. Biedenharn et al. [17-23] used the Weyl tableau techniques of construction of some vectors [17] of the Gelfand-Zetlin basis in terms of the bosons operators. The maximal and semi-maximal states of SU(n) are defined by Biedenharn et al.[18], and their importance for the study of the space of representation was observed by Moshinsky [15] and their extension to kernel and the branching operators was find in the papers of Louck [18] and Henrich [23] .

Furthermore, Nagel et Moshinsky[14] derive the Gel'fand basis polynomials in terms of the raising and lowering operators but the calculus[25-26] was very complexes and difficult to find the number of summations N of these polynomials for n>3, $N = (2^n -1) - n(n+1)/2$ [22-23]. After that, Heinrich use the kernel and the branching operators to determine the polynomials and he is unable to find it for n> 3.

In other side the Wigner coefficients of SU(3) in the canonical basis were discussed by many authors[19-42]: Moshinsky observed that the Kronecker product of k representations of SU(n) could be analyzed in terms of certain representation of SU(N),where N=k(n-1). Furthermore, a large class considered of theses coefficients, for example Biedenharn et al. using the canonical unit tensor operator method and Le blanc and Rowe use the vector coherent state theory [29-33]. The method of invariants applied by Van der Wearden and finds the generating function of 3-j symbols of SU (2). This method was generalized by Resnikoff to SU (3) and derives only the results for multiplicity free. Parakash et al. [38] uses the latest methods and the expression obtained contains 33 summations and the normalization factor is difficult to calculate.

All theses methods are very complex and the Gel'fand basis of homogenous polynomials is not found for n>3 and the Wigner coefficients with multiplicity in the canonical basis are very difficult to calculate.

To solve these important and difficult problems we proposed a simple method [39-41], the generating function method [42-45], for the calculation of Gel'fand basis polynomials, the Wigner coefficients and isoscalors factors for SU(n).

Recently the author has returned to these problems [42-44] and we applied our method to calculate the Wigner coefficients for multiplicity free. However, in this work we will do a review of this method and we focus our attention to the practical sides to do the calculations of Gel'fand basis polynomials, the Wigner coefficients and isoscalors factors with multiplicity for SU(n). .

The generalization of the generating functions of SU (2) and SU (3) to SU(n) is easy after our introduction of the binary representations of the vectors of the fundamental representations. We observe also that there is a connection between the generating function, the kernel and the branching operators expressed as functions of complex variables of SU (n). We use these functions and a recurrence method for the determination of the vectors basis of representation of SU (4). We also use the space of parameters of the generating function and the invariants method to find an algebraic expression of Wigner's coefficient in the general case, multiplicity free or not, and the isoscalors of SU (3).

This paper is organized as follows: Part two and three are a simple revision of Gel'fand basis, The fundamental representations, Matrix elements, Bosons polynomials and kernel function of SU(n). The next section is devoted to the derivation of the Generating function of SU(n). We outline the method for calculating the bosons polynomials of Gel'fand basis vector and we apply it to the case of SU(3) and SU (4) in part 6. In part 7 we present the invariant method for the calculation of Wigner's coefficients of SU(n) and we apply it to SU (2). The parts eight are devoted to the derivation of the analytic function of the 3-j symbols and the Isoscalors factors with multiplicity of SU(3). In the appendix we give a maple program very useful for the derivation of the generating function of U(n) and the normalization of Gel'fand basis.

## 2. Gel'fand basis and the fundamental representations

We summarize in this part the results of the determination of Gel'fand basis of the irreducible representation and the properties of this basis. By analogy with the theory of angular momentum, the maximal and the semi maximal of this basis are derived.
We define also the vectors of the fundamental representations.
Nagel and Moshinsky have found that the states of SU(n) may be written in terms of raising and lowering as in the theory of SU(2) and we also summarize this work.

### 2.1 The Weyl generators and the Weyl branching law of U(n)

The $n^2$ Weyl infinitesimal generators $E_{ij}$, $(i, j = 1....n)$ of the unitary group U(n) obey the commutation relations

$$[E_{ij}, E_{kl}] = \delta_{jk} E_{il} - \delta_{il} E_{kj}, \tag{2.1}$$

These generators may be written in terms of creations and destruction of n-dimensional harmonic oscillators as:

$$E_{ij} = \sum_{ij} a_i^+ a_j \tag{2.2}$$

The irreducible representations of U(n) are labeled by n-integer numbers

$$[h_{1n}, h_{2n},..., h_{nn}]. \tag{2.3}$$

When the group U(n) is restricted to the subgroup U(n-1) we find the Weyl branching law: $\quad h_{1,n} \geq h_{1,n-1} \geq h_{2,n} \geq h_{2,n-1} \geq .... \geq h_{n-1,n-1} \geq h_{n,n}$.

### 2.2 Gel'fand Basis for SU(n)

Gel'fand and al. [5] extend the Weyl branching law to U(n) and derived the individual orthogonal states of the representation, called Gel'fand basis $|(h)_n\rangle$:

$$|(h)_n\rangle = \begin{vmatrix} h_{1n} & h_{2n} & ... & h_{nn} \\ & h_{1n-1} & ... & h_{n-1n-1} \\ & & ................. \\ & h_{12} & h_{22} \\ & & h_{11} \end{vmatrix} = \begin{vmatrix} [h]_n \\ (h)_{n-1} \end{vmatrix} = \begin{vmatrix} [h]_n \\ [h]_{n-1} \\ (h)_{n-2} \end{vmatrix} ... \tag{2.4}$$

With $\qquad [h]_n = [h_{1n} h_{2n} ... h_{nn}]$

$\qquad h_{i,n} \geq h_{i,n-1} \geq h_{i+1,n}$

And $\qquad h_{i,n} \searrow \quad \nearrow h_{i+1,n} \quad SU(n)$

$\qquad\qquad\qquad h_{i,n-1} \qquad\qquad SU(n-1)$

In angular momentum and in particles physics [20] we have the notations:
for SU(2) $\qquad\qquad h_{12} = j+m, \quad h_{11} = j-m$

For SU(3) $\qquad h_{12} = I + \dfrac{Y}{2} + B, \quad h_{22} = -I + \dfrac{Y}{2} + B, h_{11} = I_3 + \dfrac{Y}{2} + B \tag{2.5}$

## 2.3 The Weyl dimension formula

The dimension of subspaces $[h_{\mu\nu}]$ is given by the Weyl formula:

$$d_{[h_{\mu\nu}]} = \left[\prod_{i<j}(p_{in} - p_{jn})\right]/[1!2!\cdots(n-1)!] \tag{2.6}$$

With $p_{in} = h_{in} + n - i$

## 2.4 The maximal and the semi maximal states

The eigenvalue of the diagonal generators $E_{ii}$ is:

$$E_{ii}|(h)_n\rangle = \omega_{in}|(h)_n\rangle, \text{ with } \omega_{in} = (\sum_{j=1}^{i}h_{j,i} - \sum_{j=1}^{i-1}h_{j,i-1}) \tag{2.7}$$

We associate to each state $|h_{\mu\nu}\rangle$ a vector or weight vector which has components

$$\omega(h) = (\omega_{1n}(h), \omega_{2n}(h), \ldots \omega_{nn}(h)).$$

A weight $\omega(h')$ is higher than a weight $\omega(h)$ if the first nonzero component in the difference $\omega(h') - \omega(h)$ is positive.

We note respectively $\left|\begin{array}{c}[h]_n \\ (max)_{n-1}\end{array}\right\rangle$ and $\left|\begin{array}{c}[h]_n \\ (min)_{n-1}\end{array}\right\rangle$ are the states that have the maximum and minimum of weight.

The vector $\left|\begin{array}{c}[h]_n \\ [h]_{n-1} \\ (max)_{n-2}\end{array}\right\rangle$ is the semi-maximal vector.

## 2.5 The fundamental representations

We can express an arbitrary irreducible representation of U(n) in terms of a set of subspace called the fundamental representations [20].
The fundamental representations of U(n) are the irreducible subspaces:

$$[1,0,\cdots,0], [1,1,\cdots,0], \cdots,[1,1,\cdots,1] \tag{2.8}$$

The dimension of the subspace $[\overbrace{1,1,1,\ldots,1}^{p},0,\ldots 0,n]$ is $C_n^p$. Then we deduce that the total number of vector bases of the fundamental representations is $2^n - 1$. And we observe that the weight vectors of these bases were expressed in terms of the binary number and it is easy to establish a correspondence between these weight vectors and the fundamentals Gel'fand basis.

We denote these fundamentals basis vectors by $|\Delta_{n,[i]}^p\rangle$, $i = 1,2\cdots,2^n - 1$.

Using the binomials formula $C_n^p = C_{n-1}^p + C_{n-1}^{p-1}$ and a symbolic program (Maple 8 see appendix1) we derive by recurrence all Gel'fand fundamental representations for n> 2 and the binary representation of the fundamental representations (B.F.R).

## 2.6 Explicit expression of Gel'fand basis vectors

Nagel and Moshinsky have found that each vector $|h_{\mu\nu}\rangle$ of the basis $[h_{\mu n}]$ may be deducted from the vector $\left|\begin{matrix}[h]_n\\(min)_{n-1}\end{matrix}\right\rangle$ or the vector $\left|\begin{matrix}[h]_n\\(max)_{n-1}\end{matrix}\right\rangle$ by applying the raising operators $R_\lambda^\mu$ or the lowering operators $L_\lambda^\mu$ and derived the explicit expressions of these operators. We write:

$$|(h)_n\rangle = N\prod_{\lambda=2}^{n}\prod_{\mu=1}^{k-1}(L_\lambda^\mu)^{L_\lambda^\mu}\left|\begin{matrix}[h]_n\\(max)_{n-1}\end{matrix}\right\rangle$$

$$= N'\prod_{\lambda=2}^{n}\prod_{\mu=1}^{k-1}(R_\lambda^\mu)^{R_\lambda^\mu}\left|\begin{matrix}[h]_n\\(min)_{n-1}\end{matrix}\right\rangle \quad (2.9)$$

With $L_\lambda^\mu = h_{\mu,\lambda} - h_{\mu,\lambda-1}\searrow$, $R_\lambda^\mu = h_{\mu,\lambda-1} - h_{\mu+1,\lambda}\nearrow$

N and N' are the constants of normalization.

It is quite clear that this result is the generalization of the well-known result of angular momentum [11]. And it is very important to mention that the computation of Gel'fand basis vectors with this formula is very difficult and complicate for n >3 [25, 26].

## 3. Matrix elements, Bosons polynomials and Kernel function of SU(n)

After the classification of elementary particles a great effort has been made to study the matrix elements of unitary groups using the Gel'fand basis and the maximal and semi-maximal cases of the D-Wigner matrix elements of SU(n) are found. The maximal and semi-maximal polynomials basis in terms of bosons operators introduced by Biedenharn et al. [17] or in term of complexes variables are used by many authors [12]. Theses polynomials are functions of minors determinants as variables and it's extension to the derivation of the kernel and the branching kernel function is found [18-23].We also give in term of bosons operators the basis of U (2) and SU (3) which are very useful later in this work.

### 3.1 The D-Wigner matrix elements of SU(n)

The application of the unitary transformation to the basis $\left|\begin{matrix}[h]_n\\(h)_n\end{matrix}\right\rangle$ is:

$$T_{U_n}\left|\begin{matrix}[h]_n\\(h)_n\end{matrix}\right\rangle = \sum_{(h')}(D_{(h'),(h)}^{[h]_n}(U_n))\left|\begin{matrix}[h]_n\\(h)_n\end{matrix}\right\rangle \quad (3.1)$$

$D_{(h'),(h)}^{[h]_n}(U_n)$ Are the elements of the matrix of SU(n).

The Gel'fand states for which $h_{rs} = h_{rn}$, $1 \leq r \leq s \leq n$, is the state of highest weight.

$$D^{[h]_n}_{(\max)(\max)}(U_n) = (\det(U_n))^{h_{n,n}} \prod_{k=1}^{n-1} (u^{(12..k)}_{(12..k)})^{h_{k,n-1}-h_{k+1,n}} \tag{3.2}$$

A special result which is immediately available from tableau techniques [18] is the so called semi-maximal case:

$$D^{[h]_n}_{([h]_{n-1})_{(\max)},(\max)}(U_n) = \frac{1}{\sqrt{N}} \prod_{k=1}^{n-1} (u^{(12..k)}_{(12..k)})^{h_{k,n-1}-h_{k+1,n}} \prod_{k=1}^{n} (u^{(12..k)}_{(12..k-1,n)})^{h_{k,n}-h_{k,n-1}} \tag{3.3}$$

$u^{(12..k)}_{(12..k)}(U_n)$ Is the minors constructed from the matrix of ($U_n$).

The normalization is:

$$N = \prod_{\substack{i<j \\ 1}}^{n} \frac{(p_{i,n-1}-p_{jn})!}{(p_{i,n}-p_{j,n}-1)!} \prod_{\substack{i<j \\ 1}}^{n-1} \frac{(p_{i,n}-p_{j,n-1}-1)!}{(p_{i,n-1}-p_{j,n-1})!} \tag{3.4}$$

**The conjugate representation**

Define the transformation

$$T_{U_n} \left| \begin{matrix} [h]_n \\ (h)_n \end{matrix} \right\rangle_c = \sum_{(h')} (D^{[h]_n}_{(h'),(h)}(U_n))^* \left| \begin{matrix} [h]_n \\ (h)_n \end{matrix} \right\rangle_c \tag{3.5}$$

The conjugate of the basis states is

$$\left| \begin{matrix} [h]_n \\ (h)_n \end{matrix} \right\rangle_c \quad \text{With} \quad \left( \left| \begin{matrix} [h]_n \\ (h)_n \end{matrix} \right\rangle_c \right)_c = \left| \begin{matrix} [h]_n \\ (h)_n \end{matrix} \right\rangle \tag{3.6}$$

## 3.2 The bosons polynomials basis of U(n)

The well known isomorphism between the spaces of Bargmann-Fock with the harmonic oscillator [34] implies that we can use one or the other of these spaces.
In this work we give the expressions of kernel and branching kernel functions in the Fock-Bargmann space because the computation in this space is very convenient.
We also give the expressions of known expressions of the bases of SU(2) and SU(3) [37].

### 3.2.1 The Fock–Bargmann space

We consider the orthonormal space of dimension n $(z_1, z_2, \cdots, z_n)$, $z_i \in C_n$ with the Gaussian measure and the scalar product is:

$$(f,g) = \int \overline{f(z)} g(z) d\mu(z) \tag{3.7}$$

With $d\mu(z) = \pi^{-n} \exp(-(z,z)) \prod_{i=1}^{n} d\operatorname{Re}(z_i) d\operatorname{Im}(z_i)$

### 3.2.2 The polynomials basis of U(n)

We consider transformation

$$\left| \begin{pmatrix} [h]_n \\ (h)_n \end{pmatrix} \right\rangle \to \Gamma\left( \begin{pmatrix} [h]_n \\ (h)_n \end{pmatrix} \right)(\Delta z) \tag{3.8}$$

In this representation the Gel'fand basis will be noted by $\Gamma\begin{pmatrix} [h]_n \\ (h)_n \end{pmatrix}(\Delta(z))$.

$\{\Gamma\left(\begin{bmatrix}[h]_n\\(h)_n\end{bmatrix}\right)(\Delta z)\}$ is an orthonormal homogenous polynomials basis of the space $B([h]_n)$

with coordinates:

$$\Delta_i^1(z) = z_i^1, \Delta_{i_1 i_2}^{12}(z) = \begin{vmatrix} z_{i_1}^1 & z_{i_1}^2 \\ z_{i_2}^1 & z_{i_2}^2 \end{vmatrix}, \cdots, \Delta_{i_1..i_k}^{1...k}(z) = \begin{vmatrix} z_{i_1}^1 & \cdots & z_{i_1}^k \\ \vdots & \vdots & \vdots \\ z_{i_k}^1 & \cdots & z_{i_k}^k \end{vmatrix}, \Delta_{12...n}^{12...n}(z) = \det(z) \qquad (3.9)$$

$\{\Delta(z)\} = \{\Delta_{i_1..i_l}^{12..l}(\Delta z), i, j = 1,\ldots,n\}$ are the minors constructed from the matrix $(z_j^i)$, $i, j = 1,\cdots,n$ by the selection of rows $1,2,\ldots,l$ and columns $i_1, i_2, \ldots, i_l$.

These coordinates are independent vectors [23-24],

And if $\delta = \text{diag}(\delta_1, \delta_2, \cdots, \delta_n)$

We have $\quad \Gamma\left(\begin{bmatrix}[h]_n\\(h)_n\end{bmatrix}\right)(\Delta(\delta z)) = \delta^{\omega_1} \delta^{\omega_2 - \omega_1} \cdots \delta^{\omega_n - \omega_{n-1}} \Gamma\left(\begin{bmatrix}[h]_n\\(h)_n\end{bmatrix}\right)(\Delta z)$

And $\omega_j = h_{1,j} + \ldots + h_{j,j}$, $\quad \langle z \| \Delta_{n,[i]}^k \rangle = \Delta_{i_1..i_k}^{1...k}(z)$ $\qquad (3.10)$

### 3.3 The kernel and the branching kernel function of SU(n)

We give only the analytical expressions of kernel function and the branching kernel functions of unitary groups [23].

### 3.3.1 The kernel function is:

$$K^n(\Delta(z), \Delta(u)) = (A_n)^{-1} \Delta^e(zu^*) = \sum_{(h)_n} \Gamma\left(\begin{bmatrix}[h]_n\\(h)_n\end{bmatrix}\right)(\Delta z) \overline{\Gamma}\left(\begin{bmatrix}[h]_n\\(h)_n\end{bmatrix}\right)(\Delta u) \qquad (3.11)$$

$$\Delta^e(z) = (\Delta_1^1(z))^{e_1} (\Delta_{12}^{12}(z))^{e_2} \cdots (\Delta_{12..n}^{12...n}(z))^{e_n}$$

$$e_i = h_{i,n} - h_{i+1,n}, i \leq n-1, \text{ and } e_n = h_{nn}$$

And $\quad A_n = (\prod_{j=1}^n (p_{jn})!) \times (\prod_{j=1}^{n-1} \prod_{k=j+1}^n (p_{j,n} - p_{k,n}))^{-1}$

### 3.3.2 The branching kernel function is:

$$R_{n-1}^n(\Delta(z), \Delta(u)) = \left[\frac{A_n}{A_{n-1}^n}\right]^{1/2} \prod_{k=1}^{n-1} (\Delta_{12..(k-1),k}^{12.....k}(z,u))^{R_n^k} \prod_{k=1}^n (\Delta_{12..(k-1),n}^{12.....n}(z,u))^{L_n^k}$$

$$= \sum_{(h)_{n-2}} \Gamma_n\begin{pmatrix}[h]_n\\ [h]_{n-1} \\ (h)_{n-2}\end{pmatrix}(\Delta z) \overline{\Gamma}_{n-1}\begin{pmatrix}[h]_{n-1}\\(h)_{n-2}\end{pmatrix}(\Delta u) \qquad (3.12)$$

With $\quad L_n^m = h_{j,n} - h_{j,n-1}$, $L_n^n = h_{n,n}$, $R_n^m = h_{j,n-1} - h_{j+1,n}$, $1 \leq j \leq n-1$

And $A_{n-1}^n = A_n \left(\prod_{i<j}(p_{in-1} - p_{jn})! \prod_{i \leq j}(p_{in} - p_{jn-1} + 1)!\right) \left/ \left(\prod_{i<j}(p_{in} - p_{jn})!(p_{in-1} - p_{jn-1})!\right)\right.^{-1}$

### 3.4 The SU(2) and SU(3) basis in terms of bosons expansion

The expressions of U (2) and SU (3) in the base of the harmonic oscillator are well known [19].

### 3.4.1 The bosons expansion of U(2)

$$\begin{pmatrix} h_{12} & h_{22} \\ & h_{11} \end{pmatrix} |0\rangle = N_2 (\Delta_{12}^{12})^{h_{22}} (\Delta_{13}^{12})^{h_{23}-h_{23}} (\Delta_1^1)^{h_{11}-h_{22}} (\Delta_2^1)^{h_{12}-h_{11}} |0\rangle \tag{3.13}$$

With
$$N_2 = \left[ \frac{(h_{12} - h_{22} + 1)!}{(h_{11} - h_{22})!(h_{12} - h_{11})!(h_{12} + 1)!(h_{22})!} \right]^{1/2}$$

### 3.4.2 The bosons expansion of U(3)

$$\begin{pmatrix} h_{13} & h_{23} & 0 \\ h_{12} & h_{22} & \\ h_{11} & & \end{pmatrix} |0\rangle = (N_3)^{-\frac{1}{2}} (\Delta_{12}^{12})^{h_{22}} (\Delta_{13}^{12})^{h_{23}-h_{23}} (\Delta_1^1)^{h_{11}-h_{23}} (\Delta_2^1)^{h_{12}-h_{11}} (\Delta_3^1)^{h_{13}-h_{12}}$$

$$\times {}_2F_1 \left( h_{22} - h_{23}, h_{11} - h_{12} \middle| h_{11} - h_{23} + 1 \middle| \frac{\Delta_1^1 \Delta_{23}^{12}}{\Delta_2^1 \Delta_{13}^{12}} \right) \tag{3.14}$$

With

$$(N_3)^{-\frac{1}{2}} = \left[ \frac{(h_{11} - h_{22})!(h_{12} - h_{23})!(h_{12} - h_{22} + 1)!(h_{13} - h_{23} + 1)!}{(h_{11} - h_{23})!(h_{12} - h_{22})!(h_{12} + 1)!(h_{22})!} \right.$$

$$\left. \times \frac{(h_{12} - h_{11})!(h_{11} - h_{23})!(h_{23} - h_{22})!}{(h_{13} - h_{22} + 1)!(h_{13} - h_{12})!} \right]^{\frac{1}{2}} \tag{3.15}$$

## 4. Generating function of SU(n)

We observe that the parameters and their powers in the generating function of the basis of SU(2) and SU(3) are linked to the raising and lowering operators and their powers, then we generalized it by an empirical way [39] to SU(n) basis. And we derive it also using the kernel function.

Our introduction of the binary fundamental representation basis (B.F.R) is very useful for calculations of the generating function and the invariance, which is connected with the complement of binary numbers [40-41].

This generating function is practical for the derivation of the invariant polynomials of SU(n) from the Gel'fand basis of unitary group SU(3(n-1)).

### 4.1 The generating function of SU(2) and SU (3)

We write only the generating functions of SU (2) and SU (3) then, we deduce simply the generating function of SU (n).

#### 4.1.1 The generating function of SU(2)

$$\sum_{h_{\mu\nu}} g_2 \varphi^2(h_{\mu\nu}, (x, y)) \Gamma_2 \begin{pmatrix} [h]_2 \\ (h)_2 \end{pmatrix} (\Delta(z)) = \exp[(\Delta_1^1 y_2^1 + \Delta_2^1 x_2^1)] \tag{4.1}$$

With $g_2 = \dfrac{1}{\sqrt{(h_{12}-h_{11})!(h_{11})!}}$ and $\varphi^2(h_{\mu\nu},(x,y)) = x^{h_{12}-h_{11}} y^{h_{11}}$

### 4.1.2 The generating function of SU(3)

The generating function of SU(3) may be written in Fock-Bargmann basis [39] in the form:

$$\sum_{h_{\mu\nu}} g_3 \varphi^3(h_{\mu\nu},(x,y)) \Gamma_3 \begin{bmatrix} [h]_3 \\ (h)_3 \end{bmatrix}(\Delta(z)) =$$
$$\exp[\Delta_{12}^{(12)} y_3^2 + (\Delta_{23}^{(12)} x_2^1 z_3^2 + \Delta_{13}^{(12)} y_2^1 z_3^2) + (\Delta_1^1 y_2^1 y_3^1 + \Delta_2^1 x_2^1 y_3^1) + \Delta_3^1 x_3^1] \quad (4.2)$$

We write
$$\phi^3(h,(x,y)) = \prod_{\ell=2}^{3} \prod_{m=1}^{\ell} \left[ (x_\ell^m)^{L_\ell^m} (y_\ell^m)^{R_\ell^m} \right] \quad (4.3)$$

We find this generating function using Schwinger's approach of angular momentum.

### 4.2 The generating function of SU(n)

The generalization of (4.2) to the generating functions of SU(n) is immediate and in the representation of Fock-Bargmann [6-7] we write

$$\sum_{h_{lm}} g_n \phi^n(h, \varphi^n(x,y)) \Gamma_n \begin{bmatrix} [h]_n \\ (h)_n \end{bmatrix}(\Delta z) = \exp\left[ \sum_k \phi_k^{n,[i]}(x,y) \Delta_{n,[i]}^k(z) \right] \quad (4.4)$$

And
$$\phi^n(h,(x,y)) = \prod_{\ell=2}^{n} \prod_{m=1}^{\ell} \left[ (x_\ell^m)^{L_\ell^m} (y_\ell^m)^{R_\ell^m} \right] \quad (4.5)$$

We will calculate $\Delta_{n,[i]}^k(z)$ by the introduction of the binary fundamental representation and then we use two simple rules for the calculation of $\phi_k^{n,[i]}(x,y)$, the constant will be calculated later.

### 4.2.1 The binary fundamental representation (B.F.R) of $\Delta_{n,[i]}^k$

We associate to each miner $\Delta_{i_1...i_l}^{12...l}$ a table of n-boxes numbered from 1 to n.

We put "one" in the boxes $i_1, i_2, ..., i_l$ and zeros elsewhere.

$$\Delta_{n,[i]}^k = \Delta_{i_1...i_k}^{12...k} = \begin{vmatrix} 1 & 2 & ... & i_1 & ... & i_k & ... & n \\ 0 & 0 & ... & 1 & ... & 1 & ... & 0 \end{vmatrix} \rangle \quad (4.6)$$

It' is very important to mention from the fact that the B.F.R. $\Delta_{n,[i]}^k$ is anti-symmetric then there are a connection between this basis and the Fock space of the second quantization hence the theory of unitary group plays an important role in physics.

### 4.2.2 Calculus of coefficients $\phi_k^{n,[i]}(x,y)$

The coefficients $\phi_k^{n,[i]}(x,y)$ may be written as product of parameters $y_\lambda^\mu = y(\lambda,\mu)$ and $x_\lambda^\mu = x(\lambda,\mu)$. We determine the indices of these parameters by using the following rules:

a- We associate to each "one" which appeared after the first zero a parameter $y(\lambda,\mu)$ whose index $\lambda$ are the number of boxes and $\mu$ the number of "one" before him, plus one.
b- We associate to each zero after the first "one" a parameter $x(\lambda,\mu)$ whose index $\lambda$ is the number of boxes and $\mu$ the number of "one" before him.

### 4.3 The generating function and the kernel function of SU(n)

We have $\quad K^n(\Delta(z),\Delta(u)) = A_n^{-1}\Delta^e(zu^*)$

In multiply by $\dfrac{A_n(h_{11})!}{e_1!e_2!\cdots e_n!}$ and by summing we find

$$\sum_{e_i}\frac{A_n(h_{11})!}{e_1!e_2!\cdots e_n!}\Delta^e(zu^*) = (\sum_k \Delta^k_{n,[i]}(z)\Delta^k_{n,[i]}(u^*))^{h_{11}}$$

Replace $\Delta^k_{n,[i]}(u^*)$ by $\phi^{n,[i]}_k(x,y)$ and summing with respect to $h_{\mu\nu}$ we find:

$$\sum_{h_{\mu\nu}}\frac{A_n(h_{11})!}{e_1!e_2!\cdots e_n!}\Delta^e(z\phi) = \exp\left[\sum_k \phi^{n,[i]}_k(x,y)\Delta^k_{n,[i]}(z)\right] \tag{4.7}$$

### 4.4 Invariance by complementary of binary numbers (R-reflexion).

We know that each binary number has a complement then we deduce that $\Delta^k_{n,[i]}(z)$ has a complement $\overline{\Delta}^k_{n,[i]}(z)$, Therefore the B.F.R. is invariant by the transformation

$$\Delta^k_{n,[i]}(z) \dashrightarrow \overline{\Delta}^k_{n,[i]}(z). \tag{4.8}$$

- For SU (2) we have the transformation $\varphi_{jm} \to (-1)^{j+m}\varphi_{j-m}$ taken into account that the complement of [0 1] is [1 0] and conversely.
- For SU (3) we also deduce the R-Conjugation of Gell-Mann (Resnikoff)

$$V^{\lambda\mu}_{(t,t_0,y)} \to (-1)^{y/2-t_0} V^{\lambda\mu}_{(t,-t_0,-y)} \tag{4.9}$$

The expression of complement $\overline{\phi}^{n,[i]}_k$ may be deduced from $\phi^{n,[i]}_k$ by changing $y(\ell,m)$ by $z(\ell,-m+\ell)$ and $z(\ell,m)$ by $z(\ell,-m+\ell)$. And it follows that the expression (4.5) is invariant by the transformation (4.8). We call this property of invariance by reflection or complementarily invariance. We also note that in the basis of U(n) the complement of $[1,1,\cdots,1]$ is $|0\rangle$ in the oscillator basis and 1 in the Fock-Bargmann space.

### 4.5 The generating functions of SU(3), U(4) and U(5)

We find simply by a direct calculation of rules a and b or using the results of the symbolic program (appendix1) the generating functions of U(4) and U(5) which are very useful for later.

#### 4.5.1 The generating function of SU(3)

We write the generating function in a manner useful for computations

$$\sum_{h_{\mu\nu}} g_3 \varphi^3(h_{\mu\nu},(x,y)\Gamma_3\begin{pmatrix}[h]_3\\(h)_3\end{pmatrix}(\Delta(z)) =$$

$$\exp[\Delta^{(12)}_{12} y^2_3 + (\Delta^{(12)}_{23} x^1_2 + \Delta^{(12)}_{13} y^1_2)z^2_3 + (\Delta^1_1 y^1_2 + \Delta^1_2 x^1_2)y^1_3 + \Delta^1_3 x^1_3]. \tag{4.10}$$

### 4.5.2 The generating function of U(4)

$$\sum_{h_4} g_4 \phi_4(h, \varphi^4(x,y)) \Gamma_4 \begin{bmatrix} [h]_4 \\ (h)_4 \end{bmatrix} (\Delta z)$$

$$= \exp \begin{bmatrix} ((\Delta^1_1 y^1_2 + \Delta^1_2 x^1_2)y^1_3 + \Delta^1_3 x^1_3)y^1_4 + \Delta^1_4 x^1_4 + \\ ((\Delta^{12}_{13} y^1_2 + \Delta^{12}_{23} x^1_2)x^2_3 + \Delta^{12}_{12} y^2_3)y^2_4 + ((\Delta^{12}_{14} y^1_2 + \Delta^{12}_{24} x^1_2)y^1_3 + \Delta^{12}_{34} x^1_3)x^2_4 + \\ ((\Delta^{123}_{134} y^1_2 + \Delta^{123}_{234} x^1_2)x^2_3 + \Delta^{123}_{124} y^2_3)x^3_4 + \Delta^{123}_{123} y^3_4 + \Delta^{1234}_{1234} y^4_4 \end{bmatrix} \tag{4.11}$$

### 4.5.3 The generating function of U(5)

$$\sum_{h_4} g_5 \phi_5(h, \varphi^5(x,y)) \Gamma_5 \begin{bmatrix} [h]_5 \\ (h)_5 \end{bmatrix} (\Delta z)$$

$$= \exp \begin{bmatrix} (((\Delta_1 y^1_2 + \Delta_2 x^1_2)y^1_3 + \Delta_3 x^1_3)y^1_4 + \Delta_4 x^1_4)y^1_5 + \Delta_5 x^1_5 + \\ ((\Delta_{15} y^1_2 + \Delta_{25} x^1_2)y^1_3 + \Delta_{35} x^1_3)y^1_4 + \Delta_{45} x^1_4)x^2_5 \\ (((\Delta_{13} y^1_2 + \Delta_{23} x^1_2)x^2_3 + \Delta_{12} y^2_3)y^2_4 + ((\Delta_{14} y^1_2 + \Delta_{24} x^1_2)y^1_3 + \Delta_{34} x^1_3)x^2_4)y^2_5 + \\ (((\Delta_{135} y^1_2 + \Delta_{235} x^1_2)x^2_3 + \Delta_{125} y^2_3)y^2_4 + ((\Delta_{145} y^1_2 + \Delta_{245} x^1_2)y^1_3 + \Delta_{345} x^1_3)x^2_4))x^3_5 + \\ ((\Delta_{1345} y^1_2 + \Delta_{2345} x^1_2)x^2_3 + \Delta_{1245} y^2_3)x^3_4 + \Delta_{1235} y^3_4)x^4_5 + \\ ((\Delta_{134} y^1_2 + \Delta_{234} x^1_2)x^2_3 + \Delta_{124} y^2_3)x^3_4 + \Delta_{123} y^3_4)y^3_5 + \Delta_{1234} y^4_5 \end{bmatrix} \tag{4.12}$$

## 5. The Gel'fand basis vectors of U(n)

We will calculate by recurrence the polynomials of the irreducible representations of SU (n) using the branching kernel function. We consider the base of U (2) as a starting point, then we presents the recurrence method and we determine the bases of the groups U (3) and U (4).

### 5.1 The Gel'fand basis of U(2).

We have $\qquad \Gamma(h_{11}) = \Delta^{h_{11}}_1 / \sqrt{(h_{11})!}$

And $\qquad \Gamma\begin{pmatrix} h_{12} & & h_{22} \\ & h_{11} & \end{pmatrix} = M^{1/2}_2 \; \Delta^{h_{12}-h_{22}}_1 \Delta^{h_{12}-h_{11}}_2 \Delta^{h_{22}}_{12} \tag{5.1}$

In the notation of angular momentum [20] we write:
$$j + m = h_{12} - h_{22}, \; j - m = h_{12} - h_{11}.$$

### 5.2.1 The recurrence method for the calculation of U (n) polynomials

By considering the product of coefficients of $y^i_n = y(n,i)$ and $x^i_n = x(n,i), i = 1, \cdots, n$ appearing in the generating function of SU (n) we find the branching kernel.
We have

$$R_{n-1}^n(\Delta(z), \varphi^{n-1}(x,y)) = \left[\frac{A_n}{A_{n-1}^n}\right]^{1/2} \times$$

$$\prod_{k=1}^{n-1}(\Delta_{12..(k-1),k}^{12....k}(z,\varphi^{n-1}))^{R_n^k} \prod_{k=1}^{n}(\Delta_{12..(k-1),n}^{12....n}(z,\varphi^{n-1}))^{L_n^k} \quad (5.2)$$

$$= \sum_{(h)_{n-2}} \Gamma_n \begin{pmatrix} [h]_n \\ [h]_{n-1} \\ (h)_{n-2} \end{pmatrix} (\Delta z) \overline{\Gamma}_{n-1} \begin{pmatrix} [h]_{n-1} \\ (h)_{n-2} \end{pmatrix} (\varphi^{n-1}(x,y)) \quad (5.3)$$

But $\overline{\Gamma}_{n-1}\begin{pmatrix}[h]_{n-1}\\(h)_{n-2}\end{pmatrix}(\varphi^{n-1}(x,y) = N_{n-1}\phi^{n-1}(h,(x,y))P_{n-1}(1)$ (5.4)

And $P_2(1) = 1$.

After identification of the two sides of (5.4) we find the polynomial representations of the irreducible of U(n)

$$\Gamma_n\begin{pmatrix}[h]_n\\(h)_{n-1}\end{pmatrix}(\Delta z) = N_n P_n(\Delta(z)), \quad N_n = \frac{\sqrt{A_n}}{N_{n-1}P_{n-1}(1)\sqrt{A_{n-1}^n}} \quad (5.5)$$

### 5.3 Calculation of $P_n(1)$

By replacing (5.4) in (5.5) we identify the results and then we do the summation for the convenience of calculations, we find the expression:

$$\prod_{k=1}^{n-1}(\Delta_{12..(k-1),k}^{12....k}(\varphi^n(a,b),\varphi^{n-1}(x,y))^{R_n^k} \prod_{k=1}^{n}(\Delta_{12..(k-1),n}^{12....n}(\varphi^n(a,b),\varphi^{n-1}(x,y))^{L_n^k} = \quad (5.6)$$

$$\sum_{(h)_{n-1}}\varphi^n(h_{\mu\nu},(a,b))P_n(1)\varphi^{n-1}(h_{\mu\nu},(x,y))$$

But $\phi^n(h,(x,y)) = \prod_{\ell=2}^{n}\prod_{m=1}^{\ell}\left[(x_\ell^m)^{L_\ell^m}(y_\ell^m)^{R_\ell^m}\right] = \phi^{n-1}(h,(x,y))\prod_{m=1}^{n}\left[(x_\ell^m)^{L_\ell^m}(y_\ell^m)^{R_\ell^m}\right]$

And $\phi^{n-1}(h,(a,b))\phi^{n-1}(h,(x,y)) = \phi^{n-1}(h,(ax,by))$

If we put $u = ax$ and $v = by$ we find after identification of the two sides of
The expression (5.6):

$$\prod_{k=1}^{n-1}(\Delta_{12..(k-1),k}^{12....k}(1,\varphi^{n-1}(u,v))^{R_n^k} \prod_{k=1}^{n}(\Delta_{12..(k-1),n}^{12....n}(1,\varphi^{n-1}(u,v))^{L_n^k} =$$

$$\sum_{(h)_{n-1}} P_n(1)\varphi^{n-1}(h_{\mu\nu},(u,v)) \quad (5.7)$$

The constants $N_n$ and $P_n(1)$ are functions of Gel'fand indices of U(n).
The expression (5.7) is very important for the computing of $P_n(1)$.

### 5.4 Calculation of $P_n(1)$ for n=3, 4, 5.

We will compute P3 (1), P4 (1) using the formula (5.7).

### 1-Calculation of $P_3(1)$

Using (5.7) we find:

$$(y_2^1 + x_2^1)^{h_{12}-h_{23}}(y_2^1 + x_2^1)^{h_{23}-h_{22}} = (y_2^1 + x_2^1)^{h_{12}-h_{22}} \tag{5.8}$$

We deduce from the above expression $P_3(1) = C_{h_{12}-h_{22}}^{h_{12}-h_{11}}$

**2- Calculation of $P_4(1)$**

We will compute $P_4(1)$ using the formula (5.7).

$$((v_2^1 + u_2^1)v_3^1 + u_3^1)^{L(4,1)+R(4,2)} \times ((v_2^1 + u_2^1)u_3^2 + v_3^2)^{L(4,2)+R(4,3)}$$
$$= \sum_{(h)_{n-1}} P_4(1)\varphi^3(h_{\mu\nu},(u,v)) \tag{5.9}$$

After development of the first member and the identification with the second member we find $P_4(1)$

$$P_4(1) = \frac{(L(4,1)+R(4,2))!}{(L(3,1))!(R(3,1))!} \frac{(L(4,2)+R(4,3))!}{(R(3,2))!(L(3,2))!} \frac{(L(3,1)+R(3,2))!}{(L(2,1))!(R(2,1))!} \tag{5.10}$$

**3- Calculation of $P_5(1)$**

We will compute $P_5(1)$ using the formula (5.7).

$$(((v_2^1 + u_2^1)v_3^1 + u_3^1)v_4^1 + u_4^1)^{L(5,1)+R(5,1)} \times$$
$$(((v_2^1 + u_2^1)u_3^2 + v_3^2)v_4^2 + ((v_2^1 + u_2^1)v_3^1 + u_3^1)u_4^2)^{L(5,2)+R(5,2)} \times \tag{5.11}$$
$$((v_2^1 + u_2^1)u_3^2 + v_3^2)u_4^3 + v_4^3)^{R(5,4)+L(5,3)} = \sum_{(h)_{n-1}} P_5(1)\varphi^4(h_{\mu\nu},(u,v))$$

After development of the first member and the identification with the second member we find $P_5(1)$

$$P_5(1) = \frac{(L(5,1)+R(5,1))!}{(R(4,1))!(L(4,1))!} \frac{(L(5,2)+R(5,2))!}{(R(4,2))!(L(4,2))!} \frac{(L(5,4)+R(5,3))!}{(R(4,3))!(L(4,3))!} \times$$
$$\frac{(L(4,2)+R(4,3))!}{(R(3,2))!(L(3,2))!} \frac{(L(4,1)+R(4,2))!}{(R(3,1))!(L(3,1))!} \frac{(L(3,1)+R(3,2))!}{(R(2,1))!(L(2,1))!} \tag{5.12}$$

## 6. The Gel'fand basis of U(3) and U(4)

We will determine the polynomials basis of SU(3) and SU(4).

**6.1 The Gel'fand basis of U(3)**

We know that $P_2(1) = 1$ so we can do the calculations with the aid of (5.5) and (5.6).
In this case, we write

$$R_2^3(\Delta(z),\varphi^2)) = \sqrt{\frac{A_2}{A_2^3}}[(\Delta_1(z)y_2^1 + \Delta_2(z)x_2^1)y_3^1]^{h_{12}-h_{23}}\Delta_3(z)^{h_{13}-h_{12}} \times$$
$$\Delta_{12}(z)^{h_{22}-h_{23}}(y_3^2)^{m_{22}}[(\Delta_{13}(z)y_2^1 + \Delta_{23}(z)x_2^1)x_3^2]^{h_{23}-h_{22}}\Delta_{123}(z)^{h_{33}} \tag{6.1}$$

Using (5.5) we find:

$$R_2^3(\Delta(z),\varphi^2)) = \sum_{h_{11}} \Gamma\binom{[h]_3}{(h)_2}(\Delta(z)) \times (N_2 \times (y_2^1 y_3^1)^{h_{12}-h_{22}}(x_2^1 y_3^1)^{h_{12}-h_{11}}(y_3^2)^{h_{22}}) \tag{6.2}$$

After identification we find the expression of the vector basis of U(3):

$$\Gamma\binom{[h]_3}{(h)_2}(\Delta(z)) = \sqrt{\frac{A_2}{A_2^3}} \sum_{i+j=h_{11}-h_{22}} \left(C_i^{h_{12}-h_{23}} \times C_j^{h_{23}-h_{22}}\right) \qquad (6.3)$$

$$\times \Delta_1^i \Delta_2^{h_{12}-h_{23}-i} \Delta_3^{h_{13}-h_{12}} \Delta_{12}^{h_{22}-h_{33}} \Delta_{13}^j \Delta_{23}^{h_{23}-h_{22}-j} \Delta_{123}^{h_{33}}$$

We find the same expression already found in paper [19, 23].

## 6.2 The Gel'fand basis of U(4)

We have

$$R_3^4(\Delta(z),\varphi) = \sum_{(h)_2} \Gamma\binom{[h]_4}{(h)_3}(\Delta(z)) \Gamma\binom{[h]_3}{(h)_2}(\varphi) \qquad (6.4)$$

This is also written in the form

$$R_3^4(\Delta(z),\varphi) = \sqrt{\frac{A_3}{A_3^4}}((\Delta_1^1 y_2^1 + \Delta_2^1 x_2^1)y_3^1 + \Delta_3^1 x_3^1)^{L_4^1} \times (\Delta_4^1)^{R_4^1} \times$$

$$\times ((\Delta_{13}^{12} y_2^1 + \Delta_{23}^{12} x_2^1)x_3^2 + \Delta_{12}^{12} y_3^2)^{L_4^2} \times ((\Delta_{14}^{12} y_2^1 + \Delta_{24}^{12} x_2^1)y_3^1 + \Delta_{34}^{12} x_3^1)^{R_4^2}$$

$$\times ((\Delta_{134}^{123} y_2^1 + \Delta_{234}^{123} x_2^1)x_3^2 + \Delta_{124}^{123} y_3^2)^{R_4^3} \times (\Delta_{123}^{123})^{L_4^3} \times (\Delta_{1234}^{1234})^{L_4^4} \qquad (6.5)$$

## b-the "bosons" polynomial of the irreducible representations of U(4)

by the development of (6.5) and using (5.5) we find the relation between the indices:

$$i + i_1 = R_3^1, \quad j + j_1 = L_3^2, \quad L_4^1 + R_4^2 - i - i_1 = L_3^1, \quad L_4^2 + R_4^3 - j - j_1 = R_3^2,$$
$$k + k_1 = L_4^1 - i, \quad \ell + \ell_1 = L_4^2 - j, \quad m + m_1 = R_4^2 - i_1, \quad n + n_1 = R_4^3 - j_1 \qquad (6.6)$$
$$k + \ell + m + n = R_2^1$$

We find that the number of indices five which is the exact number.
Finally the bosons polynomial is:

$$\Gamma_4\binom{[h]_4}{[h]_3}(\Delta(z)) = N_4 \sum_{ijklm} \frac{(L_4^1)!(L_4^2)!(R_4^2)!(R_4^3)!}{i!i_1!j!j_1!k!k_1!\ell!\ell_1!m!m_1!n!n_1!} \times \qquad (6.7)$$

$$(\Delta_1^1)^{k_1}(\Delta_2^1)^k(\Delta_3^1)^i(\Delta_4^1)^{R_4^1} \times (\Delta_{13}^{12})^{\ell_1}(\Delta_{23}^{12})^\ell(\Delta_{12}^{12})^{i_1} \times$$

$$(\Delta_{14}^{12})^{m_1}(\Delta_{24}^{12})^m(\Delta_{34}^{12})^j \times (\Delta_{134}^{123})^{n_1}(\Delta_{234}^{123})^n(\Delta_{124}^{123})^{j_1} \times (\Delta_{123}^{123})^{L_4^3} \times (\Delta_{1234}^{1234})^{L_4^4}$$

With $N_4$ is the normalization constant.
It is clear that our method is the only one who can solve this problem from the practical point of view.

## 7. The Wigner's symbols and the invariants of SU(n)

In this section we give the definition of invariant and its connection with the Wigner coefficients. By using the binary representation of invariants and the parameter space we show that our method gives the Van der Wearden's result of SU(2).

## 7.1 The Wigner's symbols

The direct product of two representations may be reduced according to the formula

$$[h^1] \otimes [h^2] = \sum (\rho)[h^3]_{(\rho)} \tag{7.1}$$

Where $(\rho)$ is the multiplicity or the number of time the representation is contained in $[h^1] \otimes [h^2]$.

With
$$\left| \begin{matrix} [h^3] \\ (h^3) \end{matrix} \right\rangle_\rho = \sum_{h^1 h^2} \left\langle \begin{matrix} [h^1] & [h^2] \\ (h^1) & (h^2) \end{matrix} \right\| \left. \begin{matrix} [h^3] \\ (h^3) \end{matrix} \right\rangle_\rho \times \left| \begin{matrix} [h^1] \\ (h^1) \end{matrix} \right\rangle \left| \begin{matrix} [h^2] \\ (h^2) \end{matrix} \right\rangle \tag{7.2}$$

The coefficients in this expression are the Clebsh-Gordan coefficients.

The vector
$$\frac{1}{\sqrt{d_{h^3}}} \sum_{(h^3)} \left| \begin{matrix} [h^3] \\ (h^3) \end{matrix} \right\rangle_\rho \left| \begin{matrix} [h^3] \\ (h^3) \end{matrix} \right\rangle_c \tag{7.3}$$

Is an invariant by unitary transformation with unity norm in the product of trois spaces. When we replace it with the above mentioned:

$$H_{(\rho)} = \sum_{h^1 h^2} \begin{pmatrix} [h^1] & [h^2] & [h^3] \\ (h^1) & (h^2) & (h^3) \end{pmatrix}_\rho \prod_{i=1}^3 \left| \begin{matrix} [h^i] \\ (h^i) \end{matrix} \right\rangle \tag{7.4}$$

The coefficients
$$\begin{pmatrix} [h^1] & [h^2] & [h^3] \\ (h^1) & (h^2) & (h^3) \end{pmatrix}_\rho = \frac{1}{\sqrt{d_{h^3}}} \left\langle \begin{matrix} [h^1] & [h^2] \\ (h^1) & (h^2) \end{matrix} \right\| \left. \begin{matrix} [h^3] \\ (h^3) \end{matrix} \right\rangle_{c\rho} \tag{7.5}$$

Are Wigner's 3j symbols of SU (n) and $\rho$ is the indices of multiplicity.
$H_{(\rho)}$ is the generalization of the Van der Wearden's invariant of the group SU(2). These invariants has the following

$$T_U^{(1,2,3)} H_{(\rho)} = H_{(\rho)}, \quad \left\langle H_{(\rho)} \middle| H_{(\rho')} \right\rangle = \delta_{(\rho),(\rho')} \tag{7.6}$$

These properties mean that the invariant polynomial is function of elementary invariants. We choose $H_{(\rho)}$ as subspace of SU(3 (n-1)) which are function of the compatible elementary invariants.

$$H_{(\rho)}(\phi^1, \phi^2, \phi^3) = \sum_{h^1 h^2} \begin{pmatrix} [h^1]_n & [h^2]_n & [h^3]_n \\ (h^1)_n & (h^2)_n & (h^3)_n \end{pmatrix}_\rho \prod_{i=1}^3 \Gamma_n \begin{pmatrix} [h^i]_n \\ (h^i)_n \end{pmatrix} (^s\phi^i) =$$

$$\Gamma_{3(n-1)} \begin{pmatrix} [h]_{3(n-1)} \\ (h)_{3(n-1)} \end{pmatrix} (^s\phi) \tag{7.7}$$

We note for the remainder of the variables by xi ($\lambda$,$\mu$), yi ($\lambda$,$\mu$), $N^i_{(3,0)}$, $P^i_{(3,0)}(1)$
Li($\lambda$,$\mu$), Ri($\lambda$, $\mu$).

## 7.2 The elementary invariants $^s\Delta^i_n(z)$ and $^s\phi^i_n$

We determine the elementary scalars $^s\Delta^i_n(z)$ which are the basic elements of the Gel'fand basis of the SU (3 (n-1)). These scalars are formed of three rows of tables, Where each row of (n-1) boxes and $\alpha_i$ "one" and zero elsewhere.
$\alpha_i$ Satisfies the following conditions

$$0 \leq \alpha_i \leq n-1, \quad \sum_{i=1}^{3} \alpha_i = n \tag{7.8}$$

## 7.3 The Wigner's coefficients of SU(2)

We will apply the formula (7.7) for the determination of 3-j symbols.

### 7.3.1 The Invariants in the Gel'fand basis

We find for SU (2) the three elementary scalars

$$\overline{|1 \quad 1 \quad 0|}, \quad \overline{|1 \quad 0 \quad 1|}, \quad \overline{|0 \quad 1 \quad 1|} \tag{7.9}$$

The parameters $\{x, y\}$ that are not in the $\{\phi_k^{3,[i]}(x,y)\}$ of elementary scalars must have the power null. We put $y_3^1 = x_3^1 = 0$ then $h_{13} = h_{23} = h_{12}$ and the invariants $H_{(\rho)}$ are the Gel'fand bases:

$$\begin{pmatrix} h_{12} & h_{12} & 0 \\ & h_{12} & h_{22} \\ & & h_{11} \end{pmatrix} \tag{7.10}$$

We can write this expression in term of well known quantum numbers of angular momentum: $h_{22} = J_3$, $h_{12} - h_{11} = J_1$, $h_{11} - h_{22} = J_2$

### 7.3.2 The elementary invariants in the space of parameters $\{{}^s\phi_3^i\}$

The elementary invariants in the space of parameters are:

$$\overline{|1 \quad 1 \quad 0|} \Rightarrow \begin{vmatrix} z_1^1 & z_2^2 \\ z_2^1 & z_2^2 \end{vmatrix} = \Delta_1^1 \Delta_2^2 - \Delta_2^1 \Delta_1^2 \Rightarrow \Xi(1,2) = \begin{vmatrix} y1(2,1) & y2(2,1) \\ x1(2,1) & x2(2,1) \end{vmatrix} \tag{7.11}$$

$$\overline{|1 \quad 0 \quad 1|} \Rightarrow \Delta_1^1 \Delta_2^3 - \Delta_2^1 \Delta_1^3 \Rightarrow \Xi(1,3) = \begin{vmatrix} y1(2,1) & y3(2,1) \\ x1(2,1) & x3(2,1) \end{vmatrix},$$

$$\overline{|0 \quad 1 \quad 1|} \Rightarrow \Delta_1^2 \Delta_2^3 - \Delta_2^2 \Delta_1^3 \Rightarrow \Xi(2,3) = \begin{vmatrix} y2(2,1) & y3(2,1) \\ x2(2,1) & x3(2,1) \end{vmatrix} \tag{7.12}$$

### 7.3.3 The generating function of 3-j symbols of SU(2)

The expression (7.7) in the case of SU (2) becomes:

$$\sum_{(h^i)_2} \begin{pmatrix} h_{12}^1 & h_{12}^2 & h_{12}^3 \\ h_{11}^1 & h_{11}^2 & h_{11}^3 \end{pmatrix} \prod_{i=1}^{3} \Gamma_2 \begin{pmatrix} h_{12}^i & 0 \\ & h_{11}^i \end{pmatrix} ({}^s\phi^i) = \Gamma_3 \begin{pmatrix} h_{12} & h_{12} & 0 \\ & h_{12} & h_{22} \\ & & h_{11} \end{pmatrix} ({}^s\phi^3) \tag{7.13}$$

We obtain the well known expression of Van der Wearden with $\rho=1$.

$$\sum_{(h^i)_2} \begin{pmatrix} h_{12}^1 & h_{12}^2 & h_{12}^3 \\ h_{11}^1 & h_{11}^2 & h_{11}^3 \end{pmatrix} \prod_{i=1}^{3} \frac{(xi(2,1))^{h_{12}^i - h_{11}^i} (yi(2,1))^{h_{11}^i}}{\sqrt{(h_{12}^i - h_{11}^i)!(h_{11}^i)!}} = \tag{7.14}$$

$$\frac{\sqrt{A_2}}{N_2\sqrt{A_2^3}} C_{h_{11}-h_{22}}^{h_{23}-h_{22}} \times (\Xi(1,2))^{h_{22}} (\Xi(1,3))^{h_{11}-h_{22}} (\Xi(2,3))^{h_{12}-h_{11}}$$

To simplify the notations we write: $u^i = (xi(2,1), yi(2,1))$.

Then we find the generating function of SU(2) or the well known Van der Wearden invariant of SU(2):

$$\sum_{m_i} \left[ \prod_{i=1}^{3} \varphi_{j_i,m_i}(u^i) \right] \begin{pmatrix} j_1 & j_2 & j_3 \\ m_1 & m_2 & m_3 \end{pmatrix} = \frac{[u^2 u^3]^{(J-2j_1)}[u^3 u^1]^{(J-2j_2)}[u^1 u^2]^{(J-2j_3)}}{\sqrt{(J+1)!(J-2j_1)!(J-2j_2)!(J-2j_3)!}} \quad (7.15)$$

We have: J=j1+j2+j3 and P1=J-2j1, P2=J-2j2, P3=J-2j3.

## 8. The 3-j symbols and the Isoscalors factors of SU(3)

We deduce that the Gel'fand pattern is reduced to 7 indices variables:
The invariants polynomials are formed from one term or monomials and function of compatible product of elementary invariant scalars.

### 8.1 The Invariants of the Gel'fand basis

We find for SU(3) seven scalar elementary compatible, which are represented by the following tables:

| 1 | 0 | 1 | 1 | 0 | 0 |, | 1 | 1 | 1 | 0 | 0 | 0 |, | 1 | 0 | 0 | 0 | 1 | 1 |

| 1 | 1 | 0 | 0 | 1 | 0 |, | 0 | 0 | 1 | 0 | 1 | 1 |, | 0 | 0 | 1 | 1 | 1 | 0 |   (8.1)

| 1 | 0 | 1 | 0 | 1 | 0 |

The parameters {x, y} that are not present in the elementary scalars $\phi_k^{n,[i]}(x,y)$ must have the power null. We find:

$$k_1 = h_{34} - h_{33}, \quad k_2 = h_{33}, \quad k_3 = h_{12} - h_{23}, \quad k_4 = h_{22} - h_{33},$$
$$k_5 = (h_{13} - h_{24}) - (h_{12} - h_{23}), \quad k_6 = h_{35} - h_{23}, \quad k_7 = (h_{23} - h_{34}) - (h_{22} - h_{33}) \quad (8.2)$$

The basis of Gel'fand for the invariants is:

$$\Gamma_6 \begin{pmatrix} h_{13} & h_{13} & h_{13} & 0 & 0 & 0 \\ & h_{13} & h_{13} & h_{24} & 0 & 0 \\ & & h_{13} & h_{24} & h_{34} & 0 \\ & & & h_{13} & h_{23} & h_{33} \\ & & & & h_{12} & h_{22} \\ & & & & & h_{12} \end{pmatrix} =^s \Gamma_6 \begin{pmatrix} [h]_6 \\ (h)_6 \end{pmatrix}$$

(8.3)

### 8.2 Calculus of the invariants in the space of parameters $^s\phi_6^i$

To determine the images of invariants in the space of parameters we write

a- | 1 | 0 | 1 | 1 | 0 | 0 | ==>

$$\begin{vmatrix} z_1^1 & z_1^3 & z_1^4 \\ z_2^1 & z_2^3 & z_2^4 \\ z_3^1 & z_3^3 & z_3^4 \end{vmatrix} = \Delta_1^1 \Delta_{23}^{56} - \Delta_2^1 \Delta_{13}^{56} + \Delta_3^1 \Delta_{12}^{56} \Rightarrow W^1 = y1(3,1)x2(3,2)\Xi(1,2) + x1(3,1)y2(3,2)$$

We apply the same method for the calculation of the image of the invariants.

b- $\overline{|1\ 1\ 1\ 0\ 0\ 0|} \Rightarrow W^2 = -y2(3,1)x1(3,2)\Xi(1,2) + x2(3,1)y1(3,2)$

c- $\overline{|1\ 0\ 0\ 0\ 1\ 1|} \Rightarrow W^3 = y1(3,1)x3(3,2)\Xi(1,3) + x1(3,1)y3(3,2)$

d- $\overline{|1\ 1\ 0\ 0\ 1\ 0|} \Rightarrow W^4 = -y3(3,1)x1(3,2)\Xi(1,3) + x3(3,1)y1(3,2)$

e- $\overline{|0\ 0\ 1\ 1\ 1\ 0|} \Rightarrow W^5 = y2(3,1)x3(3,2)\Xi(2,3) + x2(3,1)y3(3,2)$

f- $\overline{|1\ 0\ 1\ 0\ 1\ 0|} \Rightarrow W^6 = -y3(3,1)x2(3,2)\Xi(2,3) + x3(3,1)y2(3,2)$

g-

$\overline{|1\ 0\ 1\ 0\ 1\ 0|} \Rightarrow W^7 = x3(3,1)y1(3,1)y2(3,1)\Xi(1,2) - x2(3,1)y1(3,1)y3(3,1)\Xi(1,3)$

$$+ x1(3,1)y2(3,1)y3(3,1)\Xi(2,3) \qquad (8.4)$$

### 8.3 The generating function of 3-j symbols of SU(3)

The expression (7.7) is written in this case as:

$$\sum_{(h^i)_3} \begin{pmatrix} [h^1]_3 & [h^2]_3 & [h^3]_3 \\ (h^1)_3 & (h^2)_3 & (h^3)_3 \end{pmatrix}_\rho \prod_{i=1}^{3} \Gamma_3 \begin{pmatrix} [h^i]_3 \\ (h^i)_3 \end{pmatrix} ({}^s\phi^i) = N_6 \prod_{i=1}^{7} [W^i]^{k_i} \qquad (8.5)$$

The development of the second side is

$$N_6 \left( \prod_{i=1}^{7} k_i! \right) \sum_{(h^i)_3} \left( \prod_{i=1}^{15} (i_{15}!)^{-1} \right) \left[ \Xi(1,2)^{P3} \Xi(1,3)^{P2} \Xi(2,3)^{P3} \right] \times$$

$$\prod_{\ell=2}^{3} \prod_{m=2}^{\ell} \left[ (xi(\ell,m))^{L_\ell^m} (yi(\ell,m))^{R_\ell^m} \right] \qquad (8.6)$$

**a-** We have

$$\prod_{i=1}^{3} \Gamma_3 \begin{pmatrix} [h^i]_3 \\ (h^i)_3 \end{pmatrix} ({}^s\phi^i) = \left( \prod_{i=1}^{3} (N_3^i) P_3^i(1) \right) \prod_{\ell=2}^{3} \prod_{m=1}^{\ell} \left[ (xi(\ell,m))^{L_\ell^m} (yi(\ell,m))^{R_\ell^m} \right] \qquad (8.7)$$

**b-** The development of the second side of (8.5) and the identification with the first member lead to a system of equations (Appendix2). The number of indices is fifteen so we have a system of fifteen equations which has the solution:

$i_1 = R3(3,1) - L3(3,2) - P3 + i_9 + i_{11} - i_6; \quad i_2 = R2(3,2) - i_{11},$

$i_3 = R1(3,2) - i_7, \quad i_4 = R2(3,1) - P2 + i_7 - i_9 + i_6,$

$i_5 = L3(3,2) - i_9,$

$i_8 = -R2(3,1) + P2 + L1(3,2) - i_6 - i_7 + i_9, \qquad (8.8)$

$i_{10} = R3(3,2) - i_6;$

$i_{12} = L2(3,2) - i_1, \quad i_{13} = P3 - R3(3,2) - i_{11} + i_6,$

$i_{14} = P2 - i_6 - i_7, \quad i_{15} = P1 - R1(3,2) - R2(3,2) + i_7 + i_{11}.$

We have also the system $k_j = i_j + i_{(j+1)}$, $j = 1..6$, $k_7 = i_{13} + i_{14} + i_{15}$.  (8.9)

It is simple to verify that these variables are function of i7-i9 then we choose for simplicity the multiplicity ρ by: ρ = k3.
We write i6 in terms of i9:  i6 = k3-L3 (3.2) + i9.
We deduce that the number of summations is three indices: i7, i9, i11.

## 8.4 The algebraic expression of Wigner's coefficients and isoscalors of SU(4)

By replacing (8.10) and (8.6) in (8.5) and by comparison we find the algebraic expression of Wigner's coefficients, and isoscalors factors of SU(3).

$$\begin{pmatrix} [h^1]_3 & [h^2]_3 & [h^3]_3 \\ (h^1)_3 & (h^2)_3 & (h^3)_3 \end{pmatrix}_\rho = N_6 \left( \prod_{i=1}^{3} (N_3^i) P_3^i (1) \right)^{-1} \prod_{i=1}^{7} (k_i!) \times$$

$$\sum_{i_7 i_9 i_{11}} \frac{\sqrt{(P+1)! \prod_{i=1}^{3}(P-2P_i)!}}{\prod_{j=1}^{15} i_j !} \begin{pmatrix} [h^1]_2 & [h^2]_2 & [h^3]_2 \\ (h^1)_1 & (h^2)_1 & (h^3)_1 \end{pmatrix} \quad (8.10)$$

As in (7.15) we write in this case P=J and Pi=Ji.

We use the well known notations of Wigner's coefficients in terms of isoscalors, { },

And 3-j symbols of SU(2). We have:

$$\begin{pmatrix} [h^1]_3 & [h^2]_3 & [h^3]_3 \\ (h^1)_3 & (h^2)_3 & (h^3)_3 \end{pmatrix}_\rho = \begin{Bmatrix} [h^1]_3 & [h^2]_3 & [h^3]_3 \\ [h^1]_2 & [h^2]_2 & [h^3]_2 \end{Bmatrix}_\rho \begin{pmatrix} [h^1]_2 & [h^2]_2 & [h^3]_2 \\ (h^1)_1 & (h^2)_1 & (h^3)_1 \end{pmatrix} \quad (8.11)$$

We find the analytic expression of the isoscalors for the canonical basis of SU(3):

$$\begin{Bmatrix} [h^1]_3 & [h^2]_3 & [h^3]_3 \\ [h^1]_2 & [h^2]_2 & [h^3]_2 \end{Bmatrix}_\rho = N_6 \frac{\prod_{i=1}^{7}(k_i!)}{\prod_{i=1}^{3}(N_3^i) P_3^i (1)} \left( \sum_{i_7 i_9 i_{11}} \frac{\sqrt{(P+1)! \prod_{i=1}^{3}(P-2P_i)!}}{\prod_{j=1}^{15} i_j !} \right) \quad (8.12)$$

## 9. Conclusion:

a- -Our method can be extended to the other classical groups and this work is of some interest for the study of n-body problem.
b- The method of generating function that originates from a simple idea: by analogy of Dirac transformation [42] I observe in seventy that the transformation from the representation of coordinated to the oscillator basis using the generating function and the Fock- Bargmann space may be very useful. This idea has a many of applications [39-46] in quantum, Atomic, nuclear physics and in group theory.
c-Our method is also very useful in the teaching of quantum mechanics for graduate and undergraduate's students.

## Acknowledgments

I thank the Professor Andrei Postnikov and Dr. A. Breidi of Metz University (France), who sent me the papers of Bargmann-Moshinsky and Heinrich that I lost during the Lebanese war.

# 10. Appendix
**Appendix1**

The maple program for the derivation of the binary representation and it is parameters representation in the generating function and the normalization coefficients of Gel'fand polynomials basis of U(n).

```
> restart:
with(linalg):
geyz:=proc(n,m)
local lam,mu,p,z,y,dlm,dplm;
y:= array(1..n,1..n);   z:= array(1..n,1..n);
dlm:= array(1..n,1..n);  dplm:= array(1..n,1..n);
for lam from 1 to n do
for mu from 1 to n do
dlm[lam,mu]:=0;dplm[lam,mu]:=0;
od;od;
p:=1;
for lam from 1 to n-1 do
for mu from 1 to (n-lam) do
dlm[lam,mu]:=m[mu,lam]-m[mu+1,lam];
dplm[lam,mu]:=m[mu+1,lam]-m[mu,lam+1];
p:=p*((z[lam,n-mu+1]**dlm[lam,mu])*(y[lam,n-mu+1]**dplm[lam,mu]));
od;od;print("Phi of BFR" ,p);    end;
ibn:=proc(n,m)
local i,i1,j,s,bn,del;
bn:= array(1..n);w:= array(1..n);del:= array(1..n);
for j from 1 to n do
del[j]:=0;od;
bn[1]:=m[n,1];
for j from 1 to n do
s:=0;
for i from 1 to j do
s:=s+ m[n-j+1,i];
od;w[j]:=s;od;
for j from 2 to n do
bn[j]:=w[j]-w[j-1]; od;
print(" BFR", bn);
i:=0;
for j from 1 to n do
if bn[j]=1 then
i:=i+1;
del[i]:=j;fi;od;
i1:=i;print(i1,    "delta", del);     end;
#  la base de Gel'fand et la formule des binomes#
    # (n!/p!(n-p)!)=(((n-1)!/(p-1)!(n-p)!)+(n-1)!/p!(n-p-1)!)#
```

```
#SU(2)  SU(3)  SU(4)   SU(5)   SU(6)#
#===========================================#
n1:=1+3+7+15+31+63; n:=6;
nt:= array(1..n);m:= array(1..n,1..n);a:= array(1..n1,1..n,1..n);
i1:=0;
for j from 1 to n do
i1:=i1+2**(j)-1;
nt[j]:=i1; od;
n1:=nt[n];
for j from 1 to n do
for k from 1 to n do
m[j,k]:=0;   od;od;
for i from 1 to n do
for j from 1 to n do
m[i,j]:=0;od;od;
for i from 1 to n1 do
for j from 1 to n do
for k from 1 to n do
a[i,j,k]:=0;
od;od;od;
      a[2,1,1]:=1;  a[2,1,2]:=0;     a[2,2,1]:=0;  a[2,2,2]:=0;
      a[3,1,1]:=1;   a[3,1,2]:=0;    a[3,2,1]:=1;   a[3,2,2]:=0;
       a[4,1,1]:=1;   a[4,1,2]:=1;   a[4,2,1]:=1;   a[4,2,2]:=0;
                # le programme#
for i from 3 to 5 do
print("=====================================");
print("----------------","the group SU(",i,") --------------------");
print("=====================================");
i3:=nt[i-1];i4:=nt[i-2];id:=i;        print(" i3= ",i3," i4= ",i4);
          # la formule des elements ai1,1=k<=i#
for j from 1 to n do
for k from 1 to n do
m[j,k]:=0;   od;od;
for k from 1 to i do
i3:=i3+1;
for j from 1 to k do
a[i3,j,1]:=1;   od;
for k1 from 1 to k do
m[k1,1]:= a[i3,k1,1];
od;print("n=",i3,m);ibn(i,m);
geyz(i,m); od;i5:=1:
              # la formule des reccurences #
    # (i!/(j!*(i-j)!))=((i-1)!/j!(i-j-1)!)+((i-1)!/(j-1)!(i-j)!)#
    # *******************************************#
 # part 1 #   print(".........part 1........");
for j from 2 to (i-1) do
```

```
t1:=((i-1)!/((j-1)!*(i-j)!));print("part 1",t1);
for k from 1 to t1 do
i3:=i3+1:i4:=i4+1:
for k1 from 1 to (j) do
a[i3,1,k1]:=1;m[1,k1]:=a[i3,1,k1];   od;
for k2 from 2 to n do
for k3 from 1 to (n) do
a[i3,k2,k3]:= a[i4,k2-1,k3];m[k2,k3]:= a[i3,k2,k3];   od;od;
print("n=",i3,m);ibn(i,m);geyz(i,m);   od;"end k";
 # part 2 #  print(".........part 2........");
t2:=((i-1)!/(j!*(i-j-1)!));print("part 2",t2);
i5:=i4;
for k from 1 to t2 do
i3:=i3+1;i4:=i4+1;
for k1 from 1 to (j) do
a[i3,1,k1]:=1;m[1,k1]:=a[i3,1,k1];   od;
for k2 from (2) to n do
for k3 from 1 to (n) do
a[i3,k2,k3]:= a[i4,k2-1,k3];m[k2,k3]:= a[i3,k2,k3];
od;od;print("n=",i3,m);ibn(i,m);geyz(i,m);"end k";od;
i4:=i5;od;"end j";#+++++++++++++++#
           # la formule des elements aii===#
print("la formule des elements aii=======");
i3:=i3+1:i4:=i4+1:
for k1 from 1 to (id) do
for j1 from 1 to (id-k1+1) do
a[i3,k1,j1]:=1;od;od;
for k1 from 1 to (id) do
for j1 from 1 to (id-k1+1) do
m[k1,j1]:= a[i3,k1,j1]; od;od;
print("n=",i3,m);ibn(i,m);geyz(i,m);
od;"end i";> restart:
with(linalg):
   #calcul de A(m(1,n),m(1,n),...,m(n,n) de Kernel functions#
n:=3; m:= array(1..n,1..n);
coefr:=proc(n,m)
local a,mu1,mup,i,j,k,p,pp,q,qq,mq,coefn,
       coefap,n1,a1,ap,ap1;
coefn:= array(1..n);
             #part 1 Kernel functions#
ap:=1;ap1:=1;n1:=n-1;
for j from 1 to n1 do
a1:=m[j,n1]; ap:=(a1+n1-j)!*ap;
od;
for j from 1 to (n1-1) do
for k from j+1 to n1 do
```

```
 a1:=(m[j,n1]-m[k,n1]+k-j)!; ap1:=a1*ap1;
od;od;coefa:=ap1/ap;print(coefa,1);
print("***************");
     #part 2 The branching operators#
         #calcul de P( mu, mu)#p:=1;
for k from 1 to n do
for j from 1 to (k-1) do
mu1:=m[k,n]+n-k;
mup:=m[j,n]+n-j;
p:=p*((mup-mu1)!); od;od;print(p,2);
           #calcul de P( mup, mup)#
pp:=1;
for k from 1 to (n-1) do
for j from (1) to (k-1) do
 mu1:=m[k,n-1]+n-k-2;  mup:=m[j,n-1]+n-j-2;
pp:=pp*((mup-mu1)!);
od;od;print(pp,3);
           #calcul de Q( mu, mup)#q:=1;
for k from 2 to n do
mu1:=m[k,n]+n-k;
for j from 1 to (k-1) do
mup:=m[j,n-1]+n-j-1; q:=q*((mup-mu1)!);
od;od;print(q,4);
     #calcul de Q( mup, mu)#
qq:=1;
for k from 1 to n-1 do
mu1:=m[k,n-1]+n-k-1;
for j from 1 to (k) do
mup:=m[j,n]+n-j;
qq:=qq*((mup-mu1-1)!); od;od;
print(qq,5);
        #calcul de A( mup, mup)#
mq:=1;
for j from 1 to (n) do
mu1:=m[j,n]+n-j; mq:=mq*((mu1)!);
od;print(mq,6);
coefap:=(pp*p)/((mq*qq*q)); coefn[n]:=coefa*coefap;
coefb:=[(m[1,2]+1)!*(m[2,2])!*((m[1,1]-m[2,2])!)
*((m[1,2]-m[1,1])!)]/[(m[1,2]-m[2,2]+1)!];
print("coefa=",coefa); print("coefap=",coefap);
print("coefn1[n]=",coefn[n]);
coefn[n]:=coefn[n]*coefb;
print("coefb=",coefb); print("coefn[n]=",coefn[n]);
end;coefr(n,m);
```

## Appendix 2
The linear system of indices (part 8):

$$i_2 + i_6 + i_{14} + i_{15} = L1(3,1), \quad i_4 + i_8 = L1(3,2),$$
$$i_3 + i_{10} + i_{13} + i_{15} = L2(3,1), \quad i_1 + i_{12} = L2(3,2),$$
$$i_7 + i_{11} + i_{13} + i_{14} = L3(3,1), \quad i_5 + i_9 = L3(3,2),$$
$$i_1 + i_5 + i_{13} = R1(3,1), \quad i_3 + i_7 = R1(3,2),$$
$$i_4 + i_9 + i_{14} = R2(3,1), \quad i_2 + i_{11} = R2(3,2),$$
$$i_8 + i_{12} + i_{15} = R3(3,1), \quad i_6 + i_{10} = R3(3,2),$$
$$i_6 + i_7 + i_{14} = P1, \quad i_2 + i_3 + i_{15} = P2$$
$$i_{11} + i_{13} + i_{10} = P3.$$